\documentclass{article} \usepackage{amssymb,amsmath,eucal}

\begin{document}

\def\R {{\mathbb R }}
 \def\C {{\mathbb C }}
  \def\Z{{\mathbb Z}} 
  \def\H{{\mathbb H}}

\def\vr{\mathbf {wr}}

\newcommand{\sgn}{\mathop{\mathrm{sgn}}\nolimits}

\def\GL{\mathrm {GL}}

\def\U{{\rm U}}
\def\O{{\rm O}} 
\def\Sp{{\rm Sp}} 
\def\SO{{\rm SO}}

\def\frm{\frak m}

\def\ov{\overline} 
\def\phi{\varphi} 
\def\epsilon{\varepsilon}
\def\kappa{\varkappa}

\def\le{\leqslant} 
\def\ge{\geqslant}

\def\frsl{\mathfrak{sl}_2}

\def\sl{\mathrm{SL}(2,\R)}
\def\slsim{\mathrm{SL}(2,\R)^\sim}

\renewcommand{\Re}{\mathop{\rm Re}\nolimits} 
\renewcommand{\Im}{\mathop{\rm Im}\nolimits}

\newcommand{\Arg}{\mathop{\rm Arg}\nolimits}

\newcounter{sec}
 \renewcommand{\theequation}{\arabic{sec}.\arabic{equation}}
\newcounter{punct}[sec]

\newcounter{fact} \def\fact{\addtocounter{fact}{1}{\scc \arabic{fact}}}

\renewcommand{\thepunct}{\thesec.\arabic{punct}}

\def\punct{\refstepcounter{punct}{\arabic{sec}.\arabic{punct}.  }}




\def\cL{\mathcal L} 
\def\cM{\mathcal M}
\def\cH{\mathcal H}
\def\cK{\mathcal K}

\def\mn{\mathsf{M}}

\def\wt{\widetilde}

\begin{center}
{\Large\bf 
Determinantal point processes and fermionic Fock space
}

\large\sc Neretin Yuri A.

\end{center}


{\small
In this note, we construct a canonical
embedding of the space $L^2$ over a determinantal
point process to the fermionic Fock space.
Equivalently, we show that a determinantal
process is the spectral measure
for some explicit commutative group of Gaussian
operators in the fermionic Fock space. 
}

\bigskip

{\bf\large 1. Determinantal processes}

\addtocounter{sec}{1}
\setcounter{equation}{0}

\smallskip

{\bf 1.1. Determinantal processes.} 
Let $K(x,y)$ be a function of two real variables.
Consider the integral operator
$$
\cK f(x)=\int_{\R} K(x,y)f(y)\,dy
$$
in $L^2(\R)$. 

We assume that $K(x,y)$ satisfies the following
conditions

   1. $K(x,y)=\ov{K(y,x)}$, i.e. $\cK^*=\cK$.

2. $0\le \cK\le 1$, i.e.,
$$\langle \cK f,f\rangle_{L^2}\ge 0,\qquad
 \langle (1-\cK) f,f\rangle_{L^2}\ge 0$$
for all $f\in L^2(\R)$.
We call these sets by {\it configurations}
3. the function $K(x,y)$ is $C^\infty$-smooth.

\smallskip

We say that a  {\it configuration} $\omega$ is a countable
or finite subset  in $\R$ such that for each bounded segment
$[a,b]\subset\R$ the set $\omega\cap[a,b]$ is finite.
Denote by $\Omega$ the space of all  
configurations in $\R$. 

 We define a probability measure $\mu$ on $\Omega$
by the following rule.

  Let $x_1$,\dots, $x_n$ be points of $\R$.
Consider infinitisimaly small intervals
$[x_j,x_j+dx_j]$ near  these points. 
Denote by $\Xi=\Xi(x_j,dx_j)$  the following event (set):
each interval $[x_j,x_j+dx_j]$ contains a point
of $\omega\in\Omega$.
We require that the probability of the event
$\Xi=\Xi(x_j,dx_j)$ is
$$
\Bigl\{\det\limits_{1\le i\le n, 1\le j\le n}
    K(x_i,x_j)
\Bigr\}\, dx_1\dots dx_n
.$$ 

The self-consistence of this definition 
is not self-obvious, however
it is self-consistent, see the comprehensive
 Soshnikov's survey \cite{Sosh}.

We denote by $\mu(\omega)=\mu_K(\omega)$
the measure obtained in this way.

\smallskip

{\bf 1.2. Examples. References.}
The first process of this kind was discovered 
in the famous work of 
 Dyson \cite{Dys} in 1962,
in his case
$$K(x,y)=\frac{\sin(x-y)}{\pi(x-y)}$$
The corresponding process is named the 
{\it sine-process} 
(this process is one of possible limit distributions
of eigenvalues of unitary $N\times N$ unitary matrices 
as $N\to\infty$).  

Many other proceses of this kind were discovered
later, see a collection of processes having
natural origins in \cite{Sosh} 

Recently, A. Borodin and G. Olshanski
in their works  on infinite-dimensional harmonic 
analysis discovered a collection of new processes
of this kind, see \cite{BO1}-\cite{BO3}
 (but they do not satisfy the conditions
that are necessary for construction of this paper;
their kernels $K(x,y)$ are not symmetric).

Lytvynov \cite{Lyt} earlier realized
some determinantal processes as spectral measures
for quasi-free states for canonical
commutation relations 
(he considered convolution type kernels $K(x-y)$).

 In many  interesting cases
  $\cK$ is an orthogonal projector
in $L^2$.
Obviously, the Dyson sine-kernel satisfies
this condition. 

\smallskip

{\bf 1.3. Multiplicative functionals.}
Let $a(x)$ be a $C^\infty$-smooth 
function with a compact support on $\R$.
Denote by $A:L^2(\R)\to L^2(\R)$
the operator 
\begin{equation}
Af(x)=(1+a(x))f(x)
\label{umnozhenie}
.\end{equation}

We also consider the functional
$\Psi[a](\omega)$  on 
$\Omega$ defined by
$$
\Psi_a(\omega)=\prod_{x_j\in\omega} (1+a(x_j))
.$$

{\sc Proposition.}
\begin{equation}
\int_{\Omega} \Psi_a(\omega)\,d\mu(\omega)=
\det(1+\cK(A-1))
\label{integral}
.\end{equation}

\smallskip

{\sc Lemma.}
{\it The operator $\cK(A-1)$ is contained in the trace
class.}
 
{\sc Proof.} Consider the operator
$$\bigl(\cK(A-1)\bigr)^*\bigl(\cK(A-1)\bigr)=(A^*-1)\cK^2(A-1).$$
We have 
\begin{equation}
(A^*-1)\cK^2(A-1)+(A^*-1)(\cK-\cK^2)(A-1)=(A^*-1)\cK(A-1)
\label{222}
.\end{equation}
The kernel of the operator in the right hand side is 
\begin{equation}
S(x,y)=\ov{a(x)} K(x,y) a(y) 
\label{111}
\end{equation}
The kernel  $S(x,y)$ is a compactly supported
smooth function and hence it is a kernel
of an operator with rapidly decreasing singular values.

The both summands in the left-hand side of (\ref{222})
are positive operators. Hence the eigenvalues of the first summand
are rapidly decrease. \hfill $\square$

\smallskip

Hence the Fredholm determinant in (\ref{integral})
is well defined.

Obviously, our conditions for $\cK$ and $a(x)$
are surplus (for instance, $a(x)$ can be a piece-wise smooth
function with jumps), but some restrictions in this place
are necessary.

\smallskip

{\sc Lemma.} {\it Linear combinations  of functions
$\Psi_a$ are dense in $L^2(\Omega)$}

\smallskip

{\sc Proof.} Let $X_1$,\dots $X_l$ be  disjoint segments
 on $\R$.
Denote by $\Xi(X_1;\alpha_1|\dots|X_l,\alpha_l)\subset \Omega$ the following event:
$X_j$ contains $\alpha_j$ points of $\omega\in\Omega$.
Denote by $\chi_{\alpha_1,\dots,\alpha_l}(\omega)$ the indicator
functions of this set,
i.e., 
$$\chi_{\alpha_1,\dots,\alpha_l}(\omega)=
\left\{\begin{aligned}1,\qquad
\text{if $\omega\in
\Xi(X_1;\alpha_1|\dots|X_l,\alpha_l)$}
\\
0,\qquad\text{otherwise}
\end{aligned}
\right.
$$
Fix complex $r_j$.
 Let 
$$a(x)=a[ r_1,\dots,r_l  ](x)=
\left\{  
\begin{aligned} r_j-1,\qquad x\in X_j\\
                0, \qquad  x\notin \cup X_j
\end{aligned}
\right.
$$                
Then 
$$
\Psi_{a[r_1,\dots,r_l]}(\omega)=
\sum_{\alpha_1,\dots,\alpha_l}^\infty 
r_1^{\alpha_1}\dots r_l^{\alpha_l} \chi_{\alpha_1,\dots,\alpha_l}(\omega)
$$
Evidently, the functions $\chi_{\alpha_1,\dots,\alpha_l}(\omega)$ 
can be obtained as limits
of linear combinations of 
$\Psi_{a[r_1,\dots,r_l]}(\omega)$
(for instance, by differentiation in parameters $r_j$).
\hfill $\square$

\smallskip

{\bf 1.4. Sketch of proof of (\ref{integral}).}
Let us explain how to prove (\ref{integral}).
First, assume that we have not $\R$ but a finite set
$R=\{1,2,\dots,n\}$.
 Let $K=K(i,j)$ be a symmetric function
on $R\times R$ (i.e., $K$ is a $n\times n$-matrix). 

Denote by $\Omega$ the space of all the finite subsets in $R$.
Let $I\subset R$. Let $\Xi(I)\subset \Omega$ 
be set ('event') of all $J\supset I$. 
We assume that probability $p(\Xi(I))$ is
$$ 
\det\limits_{l,m\in I} K(l,m)
$$  
Let $[I]\in\Omega$ be the one-point set consisting of $I$.
Obviously, 
$$
p[I]=\sum_{J\supset I}(-1)^{|J\setminus I|} \det\limits_{l,m\in J} K(l,m) 
$$
where $|J\setminus I|$ denotes the number of elements 
in $J\setminus I$.

If $K\ge 0$, $1-K\ge 0$, then this defines a probability measure
(otherwise, probabilities can be negative).

Now let $a=a(j)$ be a function on $R$. Define a functional
on $\Omega$ by
$$
 \Psi_a(I)=\prod_{j\in I} (1+a(j))
$$
Find the mean of 
this function over $\Omega$,
\begin{multline*}
\mn \Psi_a =
\sum_{I\subset R} \Bigl\{
\prod_{j\in I} (1+a(j))
   \sum_{J\supset I}(-1)^{|J\setminus I|} \det_{l,m\in J} K(l,m)\Bigr\}
=\\=
\sum_{J\supset R}
\Bigl\{ \det_{l,m\in J} K(l,m)
\cdot \sum_{I\subset J} (-1)^{|J\setminus I|}\prod_{j\in I} (1+a(j))
\Bigr\}
=\\=
\sum_{J\subset R}
\Bigl\{ \det_{l,m\in J} K(l,m)
\cdot \prod_{j\in J} a(j)
\Bigr\}
=\det_{p,q\in R}(1+K(p,q)a(q))
\end{multline*}

The continuous case is the same, we only must
consider Fredholm determinants and follow convergence.

\smallskip

{\bf 1.5. Coherent states.} 
The formula (\ref{integral}) 
implies the following identity
\begin{equation}
\langle \Psi_a,\Psi_{\wt a}\rangle_{L^2(\Omega)}
=\det(1+\cK\bigl[A\wt A^*-1]\bigr)
\label{gramm-1}
\end{equation}

\bigskip

{\bf\large 2. Fermionic Fock space} 

\medskip

\addtocounter{sec}{1}
\setcounter{equation}{0}

{\bf 2.1. The space of semiinfinite forms.} Let $H=V\oplus W$
 be a Hilbert space. Denote by $\Pi$ the projector
to the subspace $W$.
 Let $e_0$, $e_{-1}$, $e_{-2}$,\dots
be an orthonormal basis in $V$, let $e_1$, $e_2$,\dots
be an orthonormal basis in $W$.

A {\it good basic monomial} is a product having the form
$$
e_{k_1}\wedge e_{k_2}\wedge\dots ,\qquad k_1<k_2<\dots
$$
such that $k_j=j$ starting some place.

\smallskip

{\sc Example.} The vectors $e_1\wedge e_2\wedge\dots$ and
$e_{-666}\wedge e_2\wedge e_3\dots$ are good monomials,
and $e_0\wedge e_1\wedge e_2\dots$ is a bad monomial.

\smallskip

We denote by $\bigwedge(H;\Pi)$ the Hilbert space, whose
orthonormal basis consists of good monomials
(see \cite{Ber}, \cite{FF}).

\smallskip

{\sc Remark.} The space $\bigwedge$ is the so-called
{\it space of seminfinite forms}. It is a subspace
in the fermionic Fock space. To obtain the whole
Fock space in the usual sense, we must allow
$k_j=j+\alpha$ starting some place ($\alpha$ is a constant
depending on a monomial), see \cite{Ner-book}, IV.1.

\smallskip 

{\bf 2.2. Group of symmtries of $\bigwedge$.}
Denote by $\GL(H;\Pi)$ the group of block matrices
\begin{equation}
g=\begin{pmatrix}a&b\\c&d\end{pmatrix}
:\,V\oplus W\to V\oplus W
\label{matrix}
\end{equation}
satisfying the conditions

$1^\ast$. $g$ and $g^{-1}$ are bounded operators

$2^\ast$. $b$, $c$ are Hilbert-Schmidt operators

   $3^\ast$. $d-1$ is a trace class operator.

This group acts in the space $\bigwedge(H;\Pi)$ 
by the usual change
of variables ($e_j\mapsto ge_j$).
We denote these operators by $\lambda(g)$.
Obviously,
\begin{equation}
\lambda(g_1)\lambda(g_2)=\lambda(g_1g_2)
\label{representation}
\end{equation}
If $g$ is unitary operator, then $\lambda(g)$ also is unitary.
Otherwise, $\lambda(g)$ can be unbounded,
nevertheless all the operators $\lambda(g)$ have a common
invariant domain of definiteness, see 
\cite{Ner-book}, IV. 

\smallskip

{\sc Remark.} The condition $3^\ast$ can be
omited, but we must require the Fredholm index of
the operator $d-1$ to
be zero. Under this condition the operators $\lambda(g)$
can be defined, but their definition is not obvious
(since matrix elements of $\lambda(g)$
in this case are divergent series).
Also the identity (\ref{representation}) breaks down,
and we obtain a projective (not a linear representation),
see \cite{Ner-book}, IV.4.

\smallskip

{\bf 2.3. Coherent states.} Denote by $\Upsilon$ the
{\it vacuum
vector} 
$$\Upsilon=e_1\wedge e_2\wedge e_3\wedge\dots\in \bigwedge$$
Write the matrix $g=\{g_{kl}\}\in \GL(H;\Pi)$ as a usual
infinite matrix in the basis $e_j$.
Simultaneously, we preserve the block notation

Consider the vector 
$$\lambda(g)\Upsilon=
\bigwedge\limits_{j=1}^\infty 
\Bigl(\sum g_{jk} e_k\Bigr)
$$
Obviously (see, for instance, \cite{Ner2}),
\begin{equation}
\label{coherent-2}
\langle 
\lambda(g)\Upsilon,\lambda(\wt g)\Upsilon
\rangle_{\bigwedge}
=\det\bigl(c\wt c^*+d\wt d^*\bigr)
\end{equation}
Denote by $\Pi:H\to H$ the 
orthogonal projector to the subspace $W$.
Then the last formula can be written in the form
  \begin{equation}
\label{coherent-2-bis}
\langle 
\lambda(g)\Upsilon,\lambda(\wt g)\Upsilon
\rangle_{\bigwedge}
=\det\bigl(1+\Pi (g \wt g^*-1)\bigr)
\end{equation}

\bigskip

{\bf\large 3. Correspondence. I.}

\addtocounter{sec}{1}
\setcounter{equation}{0}

\medskip

First, let the operator $\cK:L^2(\R)\to L^2(\R)$
be a projector, $\cK^2=\cK$.

\smallskip

{\bf 3.1. Multiplications by functions.}
Let  $H=L^2(\R)$.
Let $\Pi=\cK$, then $W$ is the image of $\cK$,
$V$ is the kernel of $\cK$.

Let $a(x)$ be a smooth function on $\R$ with a compact
support (as above). Let $A:L^2(\R)\to L^2(\R)$
be the same operator
 $$Af(x)=(1+a(x))f(x)$$
 as above.

{\sc Proposition.}
$A\in \GL(L^2;\cK)$.

\smallskip

{\sc Proof.} It is sufficient to show that $\cK (A-1)$,
$(A-1)\cK$ 
are  trace class operators. This was shown in 1.3.\hfill $\square$

{\sc Remark.} Let $\mathrm{Diff}$
be the group of compactly supported diffeomorphisms of 
$\R$. The operators 
$$T(q)f(x)=f(q(x))q'(x),\qquad q\in\mathrm{Diff} $$
also are contained in $G(L^2,\cK)$. Hence the group 
$\mathrm{Diff}$ also  acts
in $\bigwedge(L^2,\cK)$.

\smallskip

{\bf 3.2. Coherent states.}
Denote 
$$\phi_a=\lambda(A)\Upsilon\in \bigwedge(L^2(\R),\cK)$$
  
By (\ref{coherent-2-bis}), we have
\begin{equation}
\label{gramm-2}
\langle 
\phi_a,\phi_{\wt a}
\rangle_{\bigwedge(L^2;\cK)}
=\det\bigl(1+\cK (A\wt A^*-1)\bigr)
\end{equation}

We observe that the inner products (\ref{gramm-1}) and
(\ref{gramm-2}) coincide. 
The system $\Psi_a$ spans $L^2(\Omega)$
and hence we obtain a canonical isometric embeding
$$
U:L^2(\Omega)\to \bigwedge(L^2(\R;\cK)),
\qquad U\Psi_a=\psi_a 
$$ 

Apparently, 
this operator is one-to-one correspondence,
but I do not know a proof. 

\bigskip

{\bf\large 4. Correspondence. II.}

\addtocounter{sec}{1}
\setcounter{equation}{0}

\medskip

Now $K(x,y)$ is an arbitrary kernel satisfying
the conditions given in 1.1.

\smallskip

{\bf  4.1. Multiplications by functions.}
Consider the space $L^2(\R)\oplus L^2(\R)$.
Consider the projector $\cL$ in this space given by
$$
\cL=
\begin{pmatrix}
1-\cK&\sqrt{\cK-\cK^2}\\
\sqrt{\cK-\cK^2}&\cK
\end{pmatrix}\,\,:\,\,
 L^2(\R)\oplus L^2(\R)\to L^2(\R)\oplus L^2(\R)
$$
Let a function $a(x)$ and the operator $A:L^2(\R)\to L^2(\R)$
 be the same
as above
Denote by $g=g_a$ the operator
$$
g=
\begin{pmatrix}
1&0\\
0&A
\end{pmatrix}\,\,:\,\,
 L^2(\R)\oplus L^2(\R)\to L^2(\R)\oplus L^2(\R)
$$

{\sc Proposition.} $g\in \GL(L^2\oplus L^2,\cL)$.

\smallskip

{\sc Proof.} It is sufficient to show that
the operator $\cL (g-1)$ is a trace class operator.
For this purpose, we evaluate
$$
  (\cL (g-1))^* \cL (g-1)=(g^*-1)\cL(g-1)=
\begin{pmatrix}0&0\\0&(A^*-1)\cK(A-1)
\end{pmatrix}
$$
Again, we obtain an operator with the kernel
(\ref{111}).
\hfill $\square$

\smallskip

{\bf 4.2. Coherent states.}
Denote
$$\phi_a=\wedge(g_a)\Upsilon$$

By (\ref{coherent-2}),
\begin{multline*}
\langle \phi_a,\phi_{\wt a}\rangle_{\bigwedge(L^2\oplus L^2;\cL)}=
\det(1+\cL(g_Ag_{\wt A^*}-1))
=\\=
\det
\Bigl\{\begin{pmatrix} 1&0\\0&1\end{pmatrix}
+
\begin{pmatrix}
1-\cK&\sqrt{\cK-\cK^2}\\
\sqrt{\cK-\cK^2}&\cK
\end{pmatrix}
\begin{pmatrix} 0&0\\0&A\wt A^*-1\end{pmatrix}
\Bigr\}=\det\bigl\{1+ \cK(A\wt A^*-1)\bigr\}
\end{multline*}

Finally, we have
\begin{equation}
\label{gramm-3}
\langle \phi_a,\phi_{\wt a}\rangle_{\bigwedge(L^2\oplus L^2,\cL)}=
\det\bigl\{1+ \cK(A\wt A^*-1)\bigr\}
\end{equation}

Again, the inner products (\ref{gramm-1} and (\ref{gramm-3})
coincide. Thus we obtain a canonical isometric embedding 
$$
U:L^2(\Omega)\to \bigwedge(L^2(\R)\oplus L^2(\R),\cL),
\qquad U\Psi_a=\psi_a 
$$

{\bf Acknowledgements.}
I am grateful to G.Olshanski for explanations
of determinantal processes.
I also thank I.Derezinski for important
remarks.

{\sf Math.Phys. Group,
Institute of Theoretical and Experimental Physics,

B.Cheremushkinskaya, 25, Moscow 117259

\& University of Vienna, Math. Dept.,
Nordbergstrasse, 15, Vienna 1090, Austria}

neretin@mccme.ru


\begin{thebibliography}{cc}

\bibitem{Ber}
Berezin, F. A. 
 {\it Some remarks on the representations 
 of commutation relations.} 
 Uspehi Mat. Nauk 24 1969 no. 4 (148), 65--88.
 

\bibitem{BO1}
Borodin A., Olshanski G.I.
 {\it Point processes and the infinite symmetric group. 
 Part VI: Summary of results.}
 Preprint, available via {\tt http://de.arxiv.org/math.RT/9904010} 

\bibitem{BO2}
Borodin A., Olshanski G.I.
{\it Harmonic analysis on the infinite-dimensional unitary group and determinantal 
point processes.}
Preprirt, available via
 {\tt http://de.arxiv.org/math.RT/0109194}

\bibitem{BO3}
Borodin A., Olshanski G.I.
 Random partitions and the Gamma kernel
 {\tt http://de.arxiv.org/math-ph/0305043}

 \bibitem{Dys}
Dyson, F. J.{\it Statistical theory of the energy levels of complex systems}.
 I-III. J. Mathematical Phys. 3 1962, 140--156, 157-265, 166-175
 
\bibitem{FF}

Feigin, B. L.; Fuks, D. B. 
 {\it Skew-symmetric invariant differential
  operators on the line and Verma modules
   over the Virasoro algebra.} 
 Funktsional. Anal. i Prilozhen. 16 (1982), no. 2, 47--63.
 
\bibitem{Lyt} 
Lytvynov E.W.,
{\it Fermion and boson random point processes as particle distributions of infinite free Fermi and Bose gases of finite density}
Rev. Math.
Phys. 14 (2002), No.10, 1073-1098)


 \bibitem{Ner-book}
 Neretin, Yu. A.
 {\it Categories of symmetries and infinite-dimensional groups.} 
  London Mathematical Society Monographs.  16,
   Oxford University Press, New York, 1996
 
 \bibitem{Ner2}
 Neretin, Yu. A. 
 {\it Structures of boson and fermion Fock spaces
  in the space of symmetric functions.}
 Acta Appl. Math. 81 (2004), no. 1-3, 233--268.
 
  \bibitem{Sosh}
 Soshnikov, A. {\it Determinantal random point fields.}
  Uspekhi Mat. Nauk 55 
(2000), no. 5, 107--160; translation in Russian Math. Surveys 55 (2000),  923--975

 \end{thebibliography}
\end{document}